\documentclass{elsart}
\usepackage{amssymb}
\usepackage{amsmath}
\journal{Physics Letter A}
\begin{document}

\begin{frontmatter}

\title{Causality and Localization Operators}

\author{F. Buscemi},
\author{G. Compagno\corauthref{cor1}}
\ead{compagno@fisica.unipa.it} \corauth[cor1]{Corresponding
author}

\address{INFM, and Dipartimento di Scienze fisiche ed astronomiche  dell'Universit\`{a} di Palermo, Via Archirafi 36, 90123 Palermo,
Italy}

\begin{abstract}
The evolution of the expectation values of one and two points
scalar field operators and of positive localization operators
generated by an instantaneous point source is non local. Non
locality is attributed  either to zero point vacuum fluctuation,
or  to non local operations or to the microcausality principle
being not satisfied.
\end{abstract}

\begin{keyword}Localization;
Causality; Hegerfeldt's theorem \PACS 03.70.+k; 03.65.Pm;
03.65.-w; 03.65.Ta
\end{keyword}
\end{frontmatter}
 \section{Introduction}The problem of non locality in
quantum mechanics originates from the studies of the propagation
effects from  time varying sources of electromagnetic (e.m.) field
\cite{Ferm,Shi}. For atomic sources, it has been shown that, some
quantities as correlations of excitation between two atoms or of
the electromagnetic field are non zero  at two spacetime points
with spacelike separation \cite{Heg3,Rubin,Pow}. Instead local
quantities, as the excitation of a second atom \cite{Com2,Com3},
appear to depend causally on  the source. Recently causal behavior
of local e.m. field  operator has been obtained while in the
correlation functions is   present  a non local source independent
part, that may be attributed to the zero point field fluctuations
\cite{Ber2,Com4}. \\Other aspects of non locality are present in
the free evolution of an initially localized field configuration
\cite{Heg,Hege,Pri1}. In this case it appears that, as a
consequence of Hegerfeldt's theorem, the wavefunctions and the
average values of some positive local observables differ from zero
outside the light cone of the initially localized region
\cite{Heg,Hege}. Non local terms appear also in  the scalar
product between the state of the evolving system and the
eigenstates of local positive
operators like the Newton-Wigner's or Glauber's \cite{Pri1,Pri2,kar}.\\
Here we shall  adopt a model  previously used by Maiani and Testa
to treat the problem of causality in Q.F.T \cite{Mate}. It
consists  of  a time dependent and localized scalar source
linearly interacting with a scalar field. The use of this model
avoids any problem relative to the definition of the particle
localization due to the traverse character of the e.m field.
\cite{Coh}. Moreover   to keep the problem simple and to avoid the
question linked to the effective localization of a quantum source
\cite{Mate,Gri} we shall take the source classical. We take the
source localized in an arbitrary small region of space and
  turned on and then off after an arbitrarly small interval of time.
  While this model can in principle be treated exactly, for our
purposes we shall limit ourselves to  second order perturbation
theory in the source field coupling constant. We shall then
calculate the time evolution of the state of the system and of the
expectation values of  one,two points and localization
operators.\section{The model}In our model a classical scalar
source is linearly coupled to a scalar field $\Phi(x)$. The source
is assumed to be localized in an arbitrarily small spacetime
region around the spacetime point $y\equiv(\textbf{y},y_{0})$ so
that it is effective for an arbitrarily small time around $y_{0}$.
The Hamiltonian that describes our system is then:
\begin{equation}\label{giuk}
H=H_{0}+H_{int}
\end{equation}
with\begin{equation} H_{0} =\frac{1}{2}\int\!\!
\left(\frac{d^{3}\textbf{k}}{2\omega}\right)\!\omega\left(a^{\dag}(\textbf{k})a(\textbf{k})+a(\textbf{k})a^{\dag}(\textbf{k})\right)
\end{equation}and
\begin{equation}\!H_{int}=g\int
_{-\infty}^{+\infty}d^{3}\textbf{x}\widehat{\Phi}(x)\delta^{4}(x-y)
\end{equation}where  $g$ is the  field-source coupling constant  and $a(\textbf{k})$ and $a^{\dag}(\textbf{k})$ are
the usual annihilation and   creation  operators satisfying the
relativistic commutator rules \cite{Bjo2}:
\begin{equation}
\left[ a(\textbf{k}),a^{\dag}(\textbf{k}')\right]= 2\omega
\delta^{3}(\textbf{k}-\textbf{k}').
\end{equation}
 The field $\Phi(x)$ may be expanded in terms of the operators $a(\textbf{k})$ and $a^{\dag}(\textbf{k})$ in the standard way.
 Before the source is turned on, the field is assumed to be in its
ground state $|0\rangle$. In the following we shall use the
interaction picture. The state at time $t$ $|\Psi\rangle$ will
then be given by:
\begin{equation} \label{all}|\Psi\rangle=U(t)|0\rangle
\end{equation}where  $U(t)$ is the interaction picture time evolution operator. $U(t)$ can be easily obtained
 by integrating   the interaction picture equation of motion arising from  Hamiltonian in eq.(\ref{giuk}) and it may be shown to have the
form:\begin{equation}\label{mikko} U(t)=
\exp\left(-ig\Theta(t-y_{0})\widehat{\Phi}(y)\right)
\end{equation}with $\Theta(t)$ is the step function. Under the hypothesis of
weak coupling ($g \ll 1$), we shall expand the evolution operator
given by eq.(\ref{mikko}) up to second order in $g$. Substituting
it in eq.(\ref{all}), the explicit form of the state  up to second
order  then becomes:\begin{equation}\label{fi0}
|\Psi\rangle=|\Psi^{(0)}\rangle+|\Psi^{(1)}\rangle+|\Psi^{(2)}\rangle
\end{equation}with\begin{eqnarray}\label{fi}
&{}&\!\!|\Psi^{(0)}\rangle=|0\rangle{}  \nonumber\\
& & {}\!\! |\Psi^{(1)}\rangle =\int\!\!  d^{3}\textbf{k}\alpha(k) a^{\dag}(\textbf{k})|\,0\,\rangle{}  \nonumber\\
 & & {}\!\! |\Psi^{(2)}\rangle=1/2\Big[-\int\!\!
d^{3}\textbf{k}\alpha(k) \alpha^{\ast}(k) |0\rangle + \!\int
\!d^{3}\textbf{k} \int \! \!d^{3}\textbf{k}'\!\alpha(k) \alpha(k')
 a^{\dag}(\textbf{k})a^{\dag}\!(\textbf{k}')|0\rangle\!\Big]
\nonumber \\ \end{eqnarray}and where\begin{equation}
 \alpha(k)=-\frac{i
g\,\Theta(t-y_{o})}{(2\pi)^{3/2}} \frac{1}{(2\omega)^{1/2}}\,e^{
ik\cdot y}.
 \end{equation}The  integrals present in eq.(\ref{fi}) are regularized
by the introduction of a cut-off $\lambda$. This effectively
constraints $\textbf{k}$ to $|\textbf{k}|\leq\lambda$ and whenever
an explicit dependence on $\lambda$ is present in the matrix
elements, we shall eventually consider the limit
$\lambda\rightarrow\infty$. The state, in  the second quantized
form given in eq.(\ref{fi0}), shall be the basis for the
calculations of the expectation values we are interested in.\\
Recently, for the e.m field generated by an atomic source, it has
been examined the possibility of measuring the arrival time of a
single particle generated by the source \cite{Dam}. Therefore, as
the next step, we shall extract from the second quantized state
$|\Psi\rangle$ of eq.(\ref{fi0}), the first quantization
wavefunction $\Psi(\textbf{x})$ describing a field quantum. This
can be accomplished by projecting the one quantum component of the
state $|\Psi\rangle$, expressed in momentum space, on the one
quantum space state $|\textbf{x}\rangle= a(\textbf{x})|0\rangle$
\cite{Rom}. From eqs.(\ref{fi0}) and (\ref{fi}) we
obtain:\begin{equation}\label{silvst} \Psi(\textbf{x})=\langle 0
|a(\textbf{x})|\Psi\rangle=\langle\textbf{x}|\,\Psi^{(1)}\rangle=
g\,\Theta(t-y_{o})\Delta_{+}(\textbf{x}-\textbf{y},t-y_{0})
\end{equation}where $\Delta_{+}$ is the positive frequency  propagator $\Delta $ function,  that can be expressed by \begin{equation}[\widehat{\Phi}_{+}(x),\widehat{\Phi}_{-}(y)]=i\Delta_{+}(x-y)
\end{equation}with $\widehat{\Phi}_{+}(\widehat{\Phi}_{-})$ is the positive (negative) frequency part of the
 field operator $\widehat{\Phi}$. The explicit form of $\Delta_{+}$ \cite{Bjo2} shows that it is not zero for
 $x-y$ spacelike.  Thus the single particle component of the state generated by the
 pointlike instantaneously source, develops in a  non local way. This
result, although surprising, appears to be in agreement with
Hegerfeldt's theorem and with previous results about the  non
causal evolution of the first quantization wavefunction
\cite{Bar,rose}. We shall comment this result in the following.
\section{Expectation  values one point field operators}
We  shall now examine the expectation  values of one point
operators functions of the scalar field. As first, lets consider
directly  the expectation value of the scalar field operator
$\widehat{\Phi}(x)$ on the state,  generated by our instantaneous
pointlike source, $|\Psi\rangle$. Using eq.(\ref{fi}) we
get:\begin{equation}\label{camp} \langle
\Psi|\widehat{\Phi}(x)|\Psi\rangle=g\Delta_{ret}(x-y)
\end{equation}where $\Delta_{ret}(x)$ is the retarded  causal  propagator function \cite{Bjo2}:\begin{equation}\Delta_{ret}(x)=\Theta(x_{0})\Delta(x)
\end{equation}
 that can be expressed in terms of  $\Delta(x)$, that
 in terms of  commutator of the field is
\cite{Bjo2}:\begin{equation}[\widehat{\Phi}(x),\widehat{\Phi}(y)]=i\Delta(x-y)
\end{equation} As because $\Delta$ is zero for spacelike argument, $\Delta_{ret}$ is  therefore retarded and zero  outside the lightcone centered on the source at spacetime point $y$. Then the evolution of the expectation value of the field,
generated by the instantaneous   pointlike source on the vacuum,
 clearly shows a causal behavior. By using instead the  one point field intensity operator
$\widehat{\Phi}^{2}(x)$, we get:
\begin{equation}\label{eleo}\langle\Psi|\widehat{\Phi}^{2}(x)|\Psi\rangle=\langle
0|\widehat{\Phi}^{2}(x)|0\rangle+g^{2}\Delta_{ret}^{2}(x-y).\end{equation}This
expectation value is the sum of  two terms. The first (where the
regularization of the integrals should be be exploited) is
independent from the field-source coupling constant. It derives
from the zero-point field fluctuations  that are always present
and in fact is non zero everywhere on the  whole space-time. The
second term is source dependent and causally retarded. In it there
are contributions due all the terms, up to second order, of the
state $|\Psi\rangle$. Thus, in order to examine the causal effects
linked to the variations of the source, it is physically obvious
from eq.(\ref{eleo}) that the vacuum contribution should be
subtracted from the total expression in agreement with previous
results \cite{Com2,Com3,Ber2}. As last we shall consider as one
point operator function the  field energy density
operator:\begin{equation}\!\mathcal{H}(x)\!=\!\frac{1}{2}\!\left(\!|\nabla\widehat{\Phi}(x)|^{2}\!\!+\!\widehat{\dot{\Phi}}(x)^{2}\!+\!m^{2}\widehat{\Phi}^{2}(x)\!\right).
\end{equation}Proceeding as before, we get for its expectation value on the state
$|\Psi\rangle$ given by eq.(\ref{fi}) and up to the order of
$g^{2}$:\begin{eqnarray}\label{insost}\langle\Psi|\mathcal{H}(x)|\Psi\rangle
& =&\langle 0|\mathcal{H}(x)|0\rangle  + \frac{1}{2}g^{2}
\Bigg(\bigg(\nabla\Delta_{ret}(x-y)\bigg)^{2} + \bigg( \partial_
{t}\Delta_{ret}(x-y)\!\bigg)^{2}\nonumber\\&+&m^{2}\Delta_{ret}^{2}(x-y)\Bigg).
\end{eqnarray}In (\ref{insost}) the energy density
expectation value can be again separated in two parts. The first
represents the vacuum contribution to energy density. The second,
source dependent term,  is expressed  in terms of powers of
 $\Delta_{ret}$ and of its derivatives. Therefore it is retarded
 and zero outside the lightcone centered on the source. Thus the field energy density term coming from the source propagates causally.\\ We may also
show that  the two point field correlations function does share
the same behavior. In fact taking the average of the product
$\widehat{\Phi}(x)\widehat{\Phi}(x')$ on the state $|\Psi\rangle$,
we obtain:\begin{equation} \label{kek} \langle
\Psi|\widehat{\Phi}(x)\widehat{\Phi}(x')|\Psi\rangle= \langle
0|\widehat{\Phi}(x)\widehat{\Phi}(x')|0\rangle  +
g^{2}\Delta_{ret}(x-y)\Delta_{ret}(x'-y)\end{equation} with the
first term on the right side of (\ref{kek}) representing the field
vacuum correlations. It now depends on the separation $x-x'$ and
is moreover not zero at spacelike separation. This is just a
property of zero point correlations.  For example it is well known
that at equal time $x_{0}=x_{0}'$ the space dependence of scalar
field correlations in vacuum at large distances go as
$1/r^{2}$\cite{Bjo2}. Also the second, source dependent, term is
  not zero for spacelike intervals $x-x'$. However from its
structure in eq.(\ref{kek}) one sees that it consists of  a
product of terms, containing either $x$ or  $x'$ each causally
connected to the source at $y$. The source dependent correlations
in $x$ and $x'$, non zero at spacelike distances, are expression
of the fact that the field at each
of these points is correlated, in a causal retarded way,  to the source at  $y$. \\
Non local effects have also been  shown to arise during the free
evolution of an initially localized field configuration
\cite{Pri1,Pri2}. Recently the case has been considered where the
state, describing the field generated by a localized source, is
subjected to the action of non unitary operator that truncates
some of its parts. After proper renormalization of the state, the
expectation value of the field intensity operator has been
calculated  on it showing a non local behavior \cite{Pri2}.
\\Here we shall analogously consider the average value of the one
point operator  $\widehat{\Phi}^{2}(x)$ on the state
$|\Upsilon\rangle$ generated from $ |\Psi\rangle$ of
eq.(\ref{fi}), by the action of the number operator
$\textstyle{\widehat{N}=\int
\frac{d^{3}\textbf{k}}{2\omega}a^{\dag}(\textbf{k})a(\textbf{k})}$.
The action of $\widehat{N}$ on $|\Psi\rangle$ eliminates its zero
point part. We obtain, except to a normalization
factor:\begin{equation}\label{tramp}
\!\langle\Upsilon|\widehat{\Phi}^{2}(x)|\Upsilon\rangle
\!\propto\! g^{2}\Theta^{2}(t-y_{0})|\Delta_{+}(x-y)|^{2}.
\end{equation}It appears that  to the expectation value of eq.(\ref{tramp}) contributes  only one term. It depends  on the source
and it does  not appear to be causal, at variance with the source
dependent  part of the expectation value of the same operator
given by eq.(\ref{eleo}). Although we shall not report here the
explicit form, we must however observe the expression of
$\widehat{N}$ in terms of the field $\widehat{\Phi}$ is non local.
So,   as a matter of the fact, the action of $\widehat{N}$ on the
state $ |\Psi\rangle$ is a non local operation  that is equivalent
to the action of a non local source. This is   expected to induce
changes  over a spacelike region. As a consequence,  the
expectation values of one point operators  on the state
$|\Upsilon\rangle$ may develop  non locally in time. This  is
clearly expressed by the appearance in eq. (\ref{tramp}) of the
propagator function $\Delta_{+}$ instead of $\Delta_{ret}$. It is
of interest to observe that in eq. (\ref{tramp})  only the first
order part of the state $|\Upsilon\rangle$ contributes. Therefore
in order to study possible non local effects  it appears to be
safer   also when one considers the free evolution, to use states
generated by the action of unitary evolution operators,
representing well localized source, as the one expressed by
eq.(\ref{mikko})
\section{Expectation values of localization
operators} The study of non locality  and its connection with
causality has often been conducted by analyzing the behavior of
local operators. For example some forms of the Hegerfeldt's
theorem refer to operators which are both local and positive
\cite{Heg,Hege}. The existence of   operators satisfying both
these requirements has  however been questioned \cite{Pere}. In
particular the one point operators $\widehat{\Phi}(x)^{2}$ e
$\mathcal{H}(x)$, that we have previously used, are indeed also
positive. The non local parts, appearing in their expectation
values, are source independent and attributable  to the vacuum
 fluctuations. If we want to analyze  the local proprieties of the system,  we   must keep only the physical  relevant part and are compelled to subtract the zero point contributions. This  is
equivalent to take the operators in their normal ordered form.
Under this form  they do not  however satisfy anymore  the
positivity condition. Their use in the context of Hegerfeldt's
theorem appears   so to be unappropriate. In order to define the
concepts of localization in quantum mechanics positive operators
have previously been used. Among them second quantized form of the
Newton-Wigner(NW) operator
 position  $\widehat{\rho}_{NW}(x)$\cite{Pri1}, whose first quantized form was initially introduced to define single
particle localization \cite{NeWi}, and the Glauber operator
$\widehat{\rho}_{G}(x)$ \cite{Gla1}, initially used in quantum
optics in the context of local photon detection. Both operators
satisfy the requirement of positivity.\\We shall  now calculate
their expectation values   on the state $|\Psi\rangle$ of
eq.(\ref{fi0}). The NW operator has the form
\cite{Pri1}:\begin{equation}\label{bonjovik}
\widehat{\rho}_{NW}(x)=
  a^{\dag}_{NW}(x)a_{NW}(x)
\end{equation}where\begin{eqnarray} a_{NW}(x) & =
&\frac{1}{(2\pi)^{3/2}}\int
\frac{d^{3}\,\textbf{k}}{(2\omega)^{1/2}} \, e^{-ik\cdot x} a(\textbf{k})\nonumber \\
 a_{NW}^{\dag}(x) & =
&\frac{1}{(2\pi)^{3/2}}\int
\frac{d^{3}\,\textbf{k}}{(2\omega)^{1/2}} \, e^{ik\cdot x}
a^{\dag}(\textbf{k}).\nonumber \\
\end{eqnarray}Its  average value on the state $|\Psi\rangle$, up to second order
in the constant coupling $g$,  is:\begin{eqnarray}\label{triba1}
 \langle\Psi|\widehat{\rho}_{NW}(x)|\Psi\rangle&=&
g^{2}\Theta^{2}(t-y_{0})
\Bigg|\!\frac{1}{(2\pi)^{3}}\!\!\int\!\!\!
\frac{d^{3}\textbf{k}}{\sqrt{2\omega}}\,e^{i\big({\scriptsize
\textbf{k}\cdot(\textbf{x}-\textbf{y}})-\omega
(t-y_{0})\big)}\!\Bigg|^{2}\nonumber\\
&=&g^{2}\Theta^{2}(t-y_{o})
\Big|\psi_{\scriptsize{\textbf{y},y_{0}}}^{NW}(\textbf{x},t)\Big|^{2}
\end{eqnarray}where $\psi_{\scriptsize{\textbf{y},y_{0}}}^{NW}$ is the
NW first quantized relativistic wavefunction   for the positive
frequency state
  localized at  $\textbf{y}$ at  time $t=y_{0}$
\cite{NeWi}. The expectation value of $\widehat{\rho}_{NW}(x)$ on
the state generated by our pointlike instantaneous source, is
therefore proportional to the square  modulus of the NW
wavefunction. Using the stationary phase method, the asymptotic
expression  of  the NW wavefunction  can be  shown to be for
$|T^{2}-r^{2}|\geqslant 1$, where $T=(t-y_{0})$ and
$r=|\textbf{x}-\textbf{y}|$:
\begin{equation}\psi_{\scriptsize{\textbf{y},y_{0}}}^{NW}(\textbf{x},t)\propto\left\{\begin{array}{ll}m\sqrt{T}(T^{2}-r^{2})^{-1}\,e^{-im\sqrt{T^{2}-r^{2}}},\quad T^{2}-r^{2}>0
\\m\sqrt{T}(r^{2}-T^{2})^{-1}\,e^{-m\sqrt{r^{2}-T^{2}}},\quad
 \, T^{2}-r^{2} <0. \end{array}\right.\end{equation} $\psi_{\scriptsize{\textbf{y},y_{0}}}^{NW}$ is non zero for spacelike
 intervals, thus non local effects show up in the evolution of (\ref{triba1}). The Glauber operator for a scalar field is
\begin{equation}
\widehat{\rho}_{G}(x)=
  \widehat{\Phi}_{-}(x)\widehat{\Phi}_{+}(x).
\end{equation}
Again its expectation value on the state $|\Psi\rangle$
is:\begin{equation}\label{triba2}
\langle\Psi|\Phi_{-}(x)\Phi_{+}(x)|\Psi\rangle=g^{2}\Theta^{2}(t-y_{0})
\Delta_{+}(x-y)\Delta_{-}(x-y)
\end{equation}where $\Delta_{-}(x)=\Delta_{+}^{\ast}(x)$. Again
from the  proprieties of $\Delta_{-}(\Delta_{+})$ we see that the
expectation value   (\ref{triba2}) is also non zero outside the
lightcone centered at $y$. It is clear that the appearance of non
local effects in $\langle \widehat{\rho}_{NW}(x)\rangle $ and
$\langle \widehat{\rho}_{G}(x)\rangle $ on the state
$|\Psi\rangle$, cannot be attributed to the presence of vacuum
fluctuations. In fact the expectation  values of these observables
 on the vacuum state $|0\rangle$ is zero.
The results of eqs. (\ref{triba1}) and (\ref{triba2}) thus may
seem to show evidence of non local effects generated by the
source, while our previous results of eqs.(\ref{camp}),
(\ref{eleo}) e (\ref{insost}), obtained using    field operators,
suggest the contrary. However in studying  non locality, one
should use operators  that do not introduce by their same
definition non local effects. In particular it is a standard
requirement in relativistic quantum field theory \cite{Weiss} that
any local operators $\widehat{O}(x)$ should satisfy the principle
of microcausality. That is $\widehat{O}(x)$ must
satisfy:\begin{equation}\label{jones}
[\widehat{O}(x),\widehat{O}(y)]=0
\end{equation} for spacelike $(x-y)$ intervals.
Now using $\widehat{\rho}_{NW}(x)$ and $\widehat{\rho}_{G}(x)$ in
place of $\widehat{O}(x)$ in eq.(\ref{jones}), we
obtain:\begin{eqnarray}\label{depe}
\Big[\widehat{\rho}_{NW}(x),\widehat{\rho}_{NW}(y)\Big]=-2\!\Big\{\!\!\Big(\!\partial_{t}\Delta_{+}(x\!-\!y)\!\!\Big)\!a_{NW}(x)a_{NW}^{\dag}(y)
\!\nonumber \\
+\Big(\!\partial_{t}\Delta_{-}(x\!-\!y)\!\!\Big)\!a_{NW}(y)a_{NW}^{\dag}(x)\!\Big\}
{}  {}
\end{eqnarray}and\begin{equation}\Big[\widehat{\rho}_{G}(x),\widehat{\rho}_{G}(y)\Big]=
\Phi_{-}(x)\Phi_{+}(y)\Delta_{+}(x-y)+
\Phi_{-}(y)\Phi_{+}(x)\Delta_{-}(x-y).
\end{equation}The appearance of the function $\Delta_{+}$ and $\Delta_{-}$ and
of their derivatives makes immediately clear  that
$\widehat{\rho}_{NW}(x)$ and $\widehat{\rho}_{G}(x)$ do not
satisfy the microcausality principle. This may induce non local
effects at spacelike distances. A manifestation of this is given
by the fact that  two NW localized states at two spacetime points
$x$ and $y$, that are eigenstates of the
 NW operator $\widehat{\rho}_{NW}(x)$, are
in general  not orthogonal \cite{Fle}.
\section{Conclusions}
To investigate the non local effects that appear in the
propagation of quantum field from time varying sources, we have
used a model consisting of a quantum scalar field linearly
interacting with a classical instantaneous pointlike source. In
our model there are not present spurious effects due to the
difficulty to localize a quantum mechanical source \cite{Mate,Gri}
or to   to define single particle localized states for traverse
fields \cite{Coh}. Our  results of the expectation values of one
and two point operators are obtained using second order
perturbation theory in the field-source coupling constant. Because
of the simplicity of the model all the expectation values can be
expressed in terms of the propagator functions $\Delta s$ whose
lightcone properties are well known.
\\We have found, in agreement with previous results
\cite{Com2,Com3,Ber2}, that non locality appears both in the
expectation value of some single point operators functions of the
field and in the two point correlation function. However the non
local  terms are source independent and due to the effect of the
field zero point fluctuations.\\
Non local behavior appears also in the one particle part of the
state that evolves under the action of the source. We have shown
that the appearance of  this non local behavior    can be traced
to the fact that  the one particle wavefunction corresponds to
take a part of the complete state obtained by unitary evolution under the action of the source. This is equivalent to the action of an extended non local source.\\
At the end we have studied the expectation values  evolution  of
the second quantized Newton-Wigner and Glauber operators. Here are
present, source dependent,  non local effects. However we have
suggested that in this case the appearance of non locality can be
traced to the fact that these operators do not satisfy the
microcausality
principle.\\
In conclusion for our system  and within our approximations, as
long as we make local measurement of operators satisfying  the
microcausality principle,do not appear  non local effects, except
for the ones due to vacuum fluctuations.

\end{document}